\documentclass[conference,a4paper]{IEEEtran}

\usepackage{graphicx}      
\usepackage{cite}
\usepackage[cmex10]{amsmath}
\usepackage[utf8]{inputenc}
\usepackage{comment}
\usepackage{caption}
\usepackage{subcaption}
\usepackage{tabularx}
\usepackage[acronym]{glossaries}
\IEEEoverridecommandlockouts           
\usepackage[ruled,longend,linesnumbered]{algorithm2e}
\SetArgSty{textup}
\SetKwBlock{Loop}{Loop}{end}
\usepackage{physics}
\usepackage{lscape} 
\usepackage{xcolor}
\usepackage{array}\usepackage{makecell}
\usepackage{multirow}
\usepackage{pbox}
\usepackage{amssymb}
\usepackage{stackengine}
\usepackage[font=footnotesize]{caption}

\title{Packetized Energy Management Controller for Residential Consumers}

\author{Hafiz Majid Hussain$^*$, Ashfaq Ahmad$^\dag$, Arun~Narayanan$^*$, Pedro H. J. Nardelli$^*$, Yongheng Yang$^\dag$\\
  $^*$LUT University, Lappeenranta 53850, Finland\\
  $^\dag$Air University, Islamabad 44000,  Pakistan\\
  $^\dag$Zhejiang University, Hangzhou 310027, China\\
  Correspondence: majid.hussain@lut.fi}

\begin{document}

\maketitle
\thispagestyle{empty}
\pagestyle{empty}

\begin{abstract}
%

In this paper, we investigate the management of
energy storage control and load scheduling in scenarios considering a grid-connected photovoltaic (PV) system using packetized energy management. 
The aim is to reduce an average aggregated
system cost through the proposed \textit{packetized energy management controller} considering household energy consumption, procurement price, load scheduling delays, PV self-sufficiency via generated renewable energy and battery degradation. 
The proposed approach solves the joint optimization problem using established heuristics, namely genetic algorithm (GA), binary particle swarm optimization (BPSO), and differential evolution (DE). 
Additionally, the performances of heuristic algorithms are also
compared in terms of the effectiveness of load scheduling with
delay constraints, packetized energy transactions, and battery
degradation cost.
Case studies have been provided to demonstrate
and extensively evaluate the algorithms. 
The numerical results  show that the proposed packetized energy management controller can considerably reduce the aggregated average system cost up to  4.7\%,  5.14\%,  and  1.35\% by GA, BPSO, and DE, respectively,  while meeting the packetized energy demand and scheduling delays requirements.
\end{abstract}



\section{INTRODUCTION}

As a result of the high penetration of renewable energy resources (RERs) and modern communication technologies, power system operations have been improved considerably in terms of sustainability and economics \cite{wang2021multi}. 
The RERs have now become an alternative solution to replace fossil fuels and protect environmental concerns. Among RERs, photovoltaic (PV) energy with the storage system is the most feasible and fast-spreading technology due to storing surplus energy, improving energy efficiency, and enhancing the stability of the system. 
Though promising, energy generation from PV is stochastic in nature to time which affects the lifetime of the storage system due to the frequent charging and discharging rate
and hence, becomes less successful to manage fluctuating peak load demand (PLD) \cite{molla2019integrated}. 
Therefore, PV with a storage system may not be the simple solution for the PLD problem. 
In this regard, energy management techniques (EMTs) are the potential way to address optimally the PLD problem and reduce energy usage cost considering demand response strategies (DRS) and the exchange of surplus energy between smart homes and interconnected microgrids.

Recently, numerous EMTs have been exploited to determine economical and optimal energy allocation plans considering energy sources (like PV and grid energy) \cite{wang2021multi, molla2019integrated, li2017multiobjective, leithon2017demand} and DRS with dynamic pricings \cite { hussain2018efficient, liu2021bayesian, ahmed2017residential} subject to the various household loads, quality of service \cite {espinosa2020packetized,de2020cyber} and energy trading \cite{shafie2017stochastic, dinh2020home, lu2020home} constraints. 
For instance, the authors in \cite{liu2021bayesian,hussain2018efficient} studied energy scheduling of residential users to reduce the peak to average ratio (PAR) and minimize energy cost.  
Ahmed et al. \cite{ahmed2017residential} examined consumer behavior patterns for the prediction of future aggregated load and analyzed different user reference models, comfort, and control parameters of appliances in the context of activation delay.
Some authors \cite{ espinosa2020packetized,de2020cyber} proposed packetized energy management
(PEM) approach to address the demand of thermostatically controlled loads (TCL) and validate the quality of service (QoS).
In contrast, the authors in \cite{shafie2017stochastic, dinh2020home, lu2020home, wang2021multi, molla2019integrated, li2017multiobjective, leithon2017demand} focused on incorporating RERs together with battery storage system and management techniques. 
Shafie et al. \cite{shafie2017stochastic} investigated energy cost minimization and consumer satisfaction level in home energy management system (HEMS) under demand response programs (DRPs), while Dinh et al. \cite{ dinh2020home} conducted a study for optimizing energy consumption costs and participating in bilateral energy trading with the main (external) grid.
Similarly, other authors \cite{lu2020home , wang2021multi, molla2019integrated, li2017multiobjective} proposed an HEMS model to reduce the peak load and energy usage costs, while Leithon et al.\cite{leithon2017demand} considered joint optimization of energy scheduling at consumer and  trading for profit maximization.
%
Some of the former works \cite { hussain2018efficient, liu2021bayesian, ahmed2017residential} have addressed consumer-centric problems such as load scheduling and energy cost minimization, but they ignored the integration of RERs and bilateral energy trading.
Others \cite{espinosa2020packetized,de2020cyber} considered interesting PEM approaches, but they did not discus the role of energy retailers (\emph{i.e.,} utilities) that coexist with consumers and provide pricing mechanisms.
In subsequent works \cite{shafie2017stochastic, dinh2020home, lu2020home , wang2021multi, molla2019integrated, li2017multiobjective, leithon2017demand}, RERs with battery storage systems were incorporated in the system model, but the impacts of user inconvenience and PEM approaches have not been explicitly studied.
Further, the above mentioned works have less thoroughly investigated bilateral energy exchange between the consumer and utility (except \cite{shafie2017stochastic, lu2020home, dinh2020home}). Moreover, most of the existing works (\emph{e.g.,} \cite{hussain2018efficient, liu2021bayesian, ahmed2017residential}  \cite{shafie2017stochastic, dinh2020home, lu2020home , wang2021multi, molla2019integrated, li2017multiobjective, leithon2017demand}) have not been considered PEM approach specifically. 

In this paper,  we propose a packetized energy management controller (P-EMC) and present a joint energy scheduling and storage system management for PV system with the aim to minimize the energy packet transactions cost, load scheduling delays, and cost of the storage battery degradation. 
The battery can be charged from roof-top PV panels and an on-grid (external) power grid. For the PV system, we assume that energy generation from PV system will first serve the load, and the remaining energy is stored in the energy storage system.
We consider three types of loads and characterized them based on arrival  time, length  of  operation  time, unit energy packets demand, and maximum   allowable delay.
For  batteries, the constraints are related to the charging/discharging operation, and the resulting degradation costs.

The main contributions of the paper are:
\begin{itemize}
    \item We propose a packetized energy management controller (P-EMC) for the household loads with the characteristics such as unit energy packets (EP), the cost of EP, and scheduling of the EP. We also model the internal pricing mechanism for the EP  transactions considering the respective constraints.
    \item The internal pricing model provides a general criteria (subject to constraints) for bilateral energy trading between users and the energy packet service provider. 
    \item The proposed P-EMC solves a joint stochastic optimization problem considering well-known optimization algorithms such as genetic algorithm (GA), binary particle swarm optimization (BPSO), and differential evolution (DE).
    
\end{itemize}
\vspace{-1ex}
\section{System Model}

Consider a residential smart home connected with renewable and non-renewable energy sources, an energy storage system (\emph{i.e.}, an energy storage battery), and collection of household loads as shown in Fig. \ref{EP}. The energy generation sources include an external utility grid and a roof-top photo-voltaic (PV) system. A P-EMC is installed in the smart home to perform the following tasks: (i) communicate with the energy sources, storages and loads in the system; and (ii) devise and actuate optimal PEM schedules for the considered energy sources, storage and loads. 
\vspace{-1ex}
\subsection{Load model}
\begin{figure}
\centering
\includegraphics[width=0.8\columnwidth]{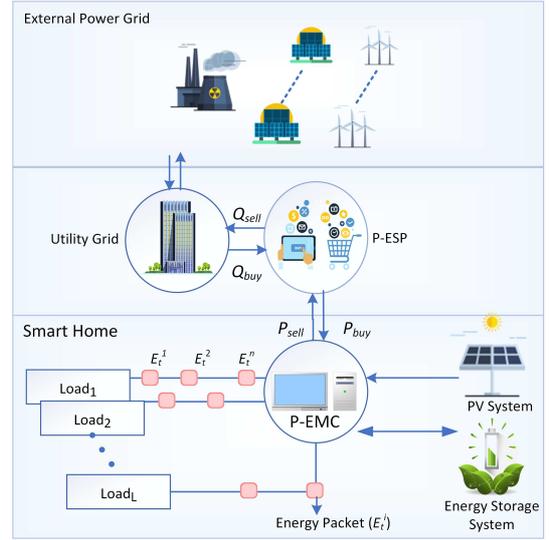}
\caption{Schematic diagram of the residential smart home}
 \label{EP}
 \vspace{-2ex}
\end{figure}

The smart home loads $i\in\{1,2,...,L\}$ are energy consumption elements that operate at discrete time slots $t\in\{0,1,2,...,T_{0}-1\}$. Each load is characterized by different attributes as follows; Load arrival time ($\lambda_{t}^{i}$), Scheduling start time ($S_{t}^{i}$), Length of operation time ($\rho_{t}^{i}$), Maximum allowable delay ($\mathrm{d}_{t,max}^i$), Load departure time ($\gamma_{t}^{i}$), and Unit energy packets demand ($E_{t}^{i}$). Consider that load $i$ consumes energy in the form of discrete value packets, and each discrete packet is denoted by $E_{t}^{i}$. Where, $E_{t}^{i}=\frac{E_{j}-E_{j-1}}{t_{k}-t_{k-1}}$, $\forall \; j\in\{1,2,...,J\}$ and $\forall \; k\in\{1,2,...,K\}$
 as illustrated in Figure \ref{EP}.

Let $n_{t}^{i}$ be the number of unit $E_{t}^{i}$ demanded by load $i$ at time slot $t$. The total energy packets demanded by all the loads ($L$) over the entire scheduling horizon ($T_{0}$) is given by the following equation.
\vspace{-2ex}
\begin{equation}
E_{T_{0}}^{L}=\sum_{i=1}^{L}\sum_{t=0}^{T_{0}-1} n_{t}^{i}\times E_{t}^{i}\label{Eq1}
\end{equation}
 Let $\mathrm d_{t}^{i}$ be the actual delay incurred by load $i$ at time slot $t$ after serving, such that,
 \vspace{-2ex}
\begin{equation}
d_{t}^{i}=\frac{S_{t}^{i}-\lambda_{t}^{i}}{d_{t,max}^{i}-\rho_{t}^{i}}\label{Eq3} 
\end{equation}
In (\ref{Eq3}), if $\lambda_{t}^{i}=S_{t}^{i}$ then $d_{t}^i$=0, and the load is immediately served. Otherwise, it is delayed as per (\ref{Eq3}). A greater value of $d_{t}^i$ in (\ref{Eq3}) means downgraded comfort level of the end-user. Thus, (\ref{Eq4}) is formulated to imposed user QoS based lower and upper limits on $d_{t}^i$.
\begin{equation}
d_{t,min}^{i}\leq d_{t}^{i}\leq d_{t,max}^{i} \label{Eq4}
\end{equation}

Following up on (\ref{Eq3}) and (\ref{Eq4}), the average experienced delay of a specific load $i$ over the entire scheduling horizon $T_0$ is calculated as follows.
\begin{equation}
\overline{d}_{T_{0}}^{i}=\frac{1}{T_{0}}\sum_{t=0}^{T_{0}-1}d_{t}^{i}, 
\label{Eq6}
\end{equation}
Finally, (\ref{Eq6}) is formulated to ensure that the user QoS based average bounds (0 and $\overline{d}_{T_0,max}^i$) on $\overline{d}_{T_0}^i$ are satisfied over the entire scheduling duration $T_0$.
\begin{equation}
0 \leq \overline{d}_{T_0}^i \leq \overline{d}_{T_0,max}^i \label{Eq6}
\end{equation}

Let $C_d(\overline{d}_{T_0}^i)$ be the function to denote the cost incurred due to $\overline{d}_{T_0}^i$ under the assumptions that $C_d(.)$ is a non-decreasing continuous convex function and its derivative ${C'}_d(.)<\infty$.  Thus, the objective here is to minimize $C_d(\overline{d}_{T_0}^i)$.   
\vspace{-1ex}
\subsection{Internal price model}

Smart home customers either have a deficiency or a surplus of energy packets. Energy deficient customers have greater energy demand than their locally generated and stored energy, and customers with an energy surplus have a smaller energy demand than their locally generated and stored energy. Energy deficient customers can buy energy packets from the external grid through an energy packet service provider (P-ESP) to meet their demand. Similarly, customers with surplus energy packets can sell them back to the external grid through the P-ESP to avoid energy wastage. 

Buying and selling of energy packets is carried through an internal pricing model of the P-ESP \cite{alizadeh2012packet}, which considers constraints of feed-in-tariff of the utility, and demand-and-supply ratio ($R_{t}^{DS}$) within the energy packet sharing zone. The P-ESP acts an agent for all the smart home prosumers. It buys energy packets from the prosumers in homes and utility grid at unit prices $P_{t}^{buy}$ and $Q_{t}^{buy}$, and sells energy packets to them at unit prices $P_{t}^{sell}$ and $Q_{t}^{sell}$, respectively. 
\begin{eqnarray}
 P_{t}^{sell} = \left\{
  \begin{array}{l l}
    \frac{Q_{t}^{sell}Q_{t}^{buy}}{(Q_{t}^{buy}-Q_{t}^{sell})R_{t}^{DS}+Q_{t}^{sell}} & \quad \text{\emph{if} $0\leq R_{t}^{DS}\leq 1$} \label{Eq7} \\
    Q_{t}^{sell} & \quad \text{\emph{otherwise}}
  \end{array} \right.
\end{eqnarray}
It is evident from (\ref{Eq7}) that: (i) if $R_{t}^{DS}=0$, the smart home prosumers do not sell energy packets and the required number of energy packets are procured from the utility at $Q_{t}^{buy}$; (ii) if $R_{t}^{DS}\geq 1$, the smart home prosumer has an energy packet surplus and this surplus is fed back to the utility at $Q_{t}^{sell}$; and (iii) if $0< R_{t}^{DS}< 1$, the selling price is dynamically adjusted between $Q_{t}^{sell}$ and $Q_{t}^{buy}$. On the other hand, internal energy packet buying price is defined in (\ref{Eq8}) considering internal energy packet selling cost, P-ESP's charge and utility's charge. 

\begin{eqnarray}
\small 
 P_{t}^{buy} = \left\{ \mkern-11mu
  \begin{array}{l r}
   P_{t}^{sell}R_{t}^{DS}+Q_{t}^{buy}(1-R_{t}^{DS}) & \mkern-8mu \text{\emph{if} $0\leq R_{t}^{DS}\leq 1$} \label{Eq8} \\
    Q_{t}^{sell} & \mkern-8mu \text{\emph{otherwise}}
  \end{array} \right.
\end{eqnarray}

In (\ref{Eq8}), $0< R_{t}^{DS}< 1$ means that the total energy packet demand is greater than the total energy packet supply of the smart home prosumers in the energy packet sharing zone, and this energy packet deficiency is fulfilled by buying energy packets from the utility at $Q_{t}^{buy}$.

%
%

Based on the load and the internal price models in this article, the cost of buying and selling energy packets from and to the utility at time slot $t$ via P-ESP can be expressed by (\ref{Eq9}) and (\ref{Eq10}), respectively, as follows.   
\begin{equation}
C_{t}^{buy}=P_{t}^{sell} \Big( E_{t}^{L}-(E_{t}^{pv}+E_{t}^{s})\Big)\;\; if\; E_{t}^{L}>E_{t}^{pv}+E_{t}^{s}\label{Eq9}
\end{equation}
\begin{equation}
C_{t}^{sell}=P_{t}^{buy} \Big((E_{t}^{pv}+E_{t}^{s})-E_{t}^{L}\Big)\;\; if\; E_{t}^{pv}+E_{t}^{s}>E_{t}^{L}\label{Eq10}
\end{equation}
Thus, the average cost of energy packets transactions ($\overline{C}_{t}^{tx}$) can be calculated by the following equation. 
\begin{equation}
\overline{C}_{t}^{tx}=\frac{1}{T_{0}}\sum_{t=0}^{T_{0}-1}\Big(C_{t}^{sell}-C_{t}^{buy}\Big)\label{Eq11}
\end{equation}

The objective is to maximize the prosumer's energy packet revenue by minimizing the difference between total energy packets selling and buying. However, this buying and selling of energy packets is constrained by the following equations:
\begin{eqnarray}
& \sum\limits_{i=1}^{L}\sum\limits_{t=0}^{T_{0}-1}x_{t}^{i}=E_{T_{0}}^{L}\label{Eq12}\\
& E_{t,min}^{i}\leq x_{t}^{i} \leq E_{t,max}^{i} \label{Eq13}\\
& E_{t}^{i}-x_{t}^{i}\leq B_{max} \label{Eq14}
\end{eqnarray}
In the above, (\ref{Eq12}) implies that flexible loads can be scheduled to operate at other allowable time slots ($x_{t}^{i}$); however, in doing so, the total energy packet demand must be kept constant. 

Similarly, (\ref{Eq13}) ensures that the scheduling of flexible loads ($x_{t}^{i}$) should not violate user's base energy packet demand ($E_{t,min}^{i}$) and the upper bound of supply capacity ($E_{t,max}^{i}$), Finally, (\ref{Eq14}) imposes a constraint on the feed-in energy packets when the utility prohibits selling of additionally generated energy packets ($B_{max}$) due to grid security issues.

\subsection{PV system}
\vspace{-1ex}
The smart home prosumers are equipped with roof-top PV panels generating renewable energy. Adopting the model in \cite{shirazi2015optimal}, let $E_{t}^{pv}$ be the amount of harvested energy from the PV source at $t$, such that,  
\begin{equation}
E_{t}^{pv}= \eta_{pv}\times A_{pv}\times I_{ir} (1-0.005(T_a((t)-25))
 \label{Eq15}   
\end{equation}
where $\eta_{pv}$ is conversion efficiency of the PV system
$A_{pv}$ is the area of the generator,
$I_{ir}$ is the solar irradiance at time t, 0.005 is temperature correction factor and $T_a$ is the outdoor temperature. We assume that $E_{t}^{pv}$ is firstly given to the scheduled load at $t$ ($x_{t}^{L}=\sum\limits_{i=1}^{L}x_{t}^{i}$), and the remaining ($r_{t}^{pv}$), if any, is stored in the energy storage system. Let the consumed portion of $E_{t}^{pv}$ be $c_{t}^{pv}$, such that,
\begin{eqnarray}
c_{t}^{pv}=\text{min}\Big\{x_{t}^{L},E_{t}^{pv}\Big\}\label{Eq16} \\
0\leq r_{t}^{pv} \leq E_{t}^{pv}-c_{t}^{pv} \label{Eq17}
\end{eqnarray}
It is worth noting here that charging and discharging activities of energy storage battery incur a degradation cost in it. Thus, the decision to store the unused portion of $E_{t}^{pv}$ (i.e., $r_{t}^{pv}$) in the battery is taken by the packetized energy controller installed in the smart home. 

\subsection{Energy storage system}
Depending on the current energy packet demand and supply conditions, energy storage system can be characterized by three possible states: charging, discharging and idle. For instance, it can be charged from a roof-top PV system, or a P-ESP or a combination of both. Similarly, it can be discharged to meet the energy packet requirement of different loads. In an idle state, it is neither charging nor discharging. These state transitions are bounded by the following set of constraints.
\begin{eqnarray}
& 0\leq r_{t}^{pv}+E_{t}^{g}\leq H_{max}\label{Eq18} \\
& 0\leq k_{t} \leq K_{max} \label{Eq19}\\
& E_{min}^{s}\leq E_{t}^{s} \leq E_{max}^{s} \label{Eq20}
\end{eqnarray}
Specifically, (\ref{Eq18}) ensures that the total charging amount at time slot $t$ ($r_{t}^{pv}+E_{t}^{g}$) does not exceed its upper bound ($H_{max}$). While, (\ref{Eq19}) limits the total discharging amount at $t$ ($k_{t}$) by its upper bound ($K_{max}$), and (\ref{Eq20}) imposes minimum and maximum capacity constraints ($E_{min}^{s}$ and $E_{max}^{s}$) on the current energy state of the battery ($E_{t}^{s}$). The dynamics of the current energy state of the storage battery evolve according to the following equation.
\begin{equation}
E_{t+1}^{s}= \alpha_{t}E_{t}^{s}+ \eta_{t}^{(+)}\Big(r_{t}^{pv}+E_{t}^{g}\Big)-\eta_{t}^{(-)}\Big(k_{t}\Big)\label{Eq21}   
\end{equation}
In (\ref{Eq21}), $\alpha_{t}$ accounts for a decay rate in the battery with the passage of time, and $\eta_{t}^{(+)}$ and $\eta_{t}^{(-)}$ denote the charging and discharging efficiencies, respectively. Let $a^{(+)}_{t} \triangleq \{\text{1,  if }r_{t}^{pv}+E_{t}^{g}>0; \text{0,  otherwise} \}$ indicate whether a charging activity occurred ($a^{(+)}_{t}=1$) or not ($a^{(+)}_{t}=0$). Similarly, $a^{(-)}_{t} \triangleq \{\text{1,  if }k_{t}>0; \text{0,  otherwise} \}$ is defined to track the occurrence of a discharging activity. These charging and discharging activities incur degradation cost in the battery, denoted by $c^{(+)}_{t}$ and $c^{(-)}_{t}$, respectively. Based on extensive analyses of the authors in \cite{ahmad2020real}, the degradation costs in the storage battery at $t$ can be modelled as follows.
\begin{eqnarray}
& c^{(+)}_{t}=\frac{h_{r}}{h_{t}} \bigg\{ \bigg(\frac{r_{r}^{pv}+E_{r}^{g}}{r_{t}^{pv}+E_{t}^{g}}\bigg)^{w_{0}}\times \exp^{w_{1}(\frac{r_{t}^{pv}+E_{t}^{g}}{r_{r}^{pv}+E_{r}^{g}}-1)}  \bigg\}\label{Eq22} \\
& c^{(-)}_{t}= \frac{h_{r}}{h_{t}} \bigg\{ \bigg(\frac{k_r}{k_t}\bigg)^{w_{2}}\times \exp^{w_{3}(\frac{k_t}{k_r}-1)}  \bigg\} \label{Eq23}
\end{eqnarray}
In (\ref{Eq22}) and (\ref{Eq23}), if the actual cyclic depth-of-charge (i.e., $r_{t}^{pv}+E_{t}^{g}$) and depth-of-discharge ($k_{t}$) are kept at their rated values (i.e., $r_{r}^{pv}+E_{r}^{g}$ and $k_{r}$), then the lifetime of the storage battery is affected by current variations corresponding to their rated values. Thus, from (\ref{Eq22}) and (\ref{Eq23}), the battery degradation cost is modelled at $t$ is given in (\ref{Eq24}) and its average over the $T_0$ duration is given in (\ref{Eq25}).
\begin{eqnarray}
& C_{t}^{s}=a_{t}^{(+)}c_{t}^{(+)}+a_{t}^{(-)}c_{t}^{(-)}  \label{Eq24} \\
& \overline{C}_{T_{0}}^{s}=\frac{1}{T_0}\sum\limits_{t=0}^{T_0-1}C_{t}^{s} \label{Eq25} 
\end{eqnarray}
Our aim is to minimize the average degradation cost in (\ref{Eq25}).
\vspace{-2ex}
\section{ Problem Formulation } \label{Problem_For}
Let ${\theta_{t}} \triangleq [E_{t}^{g},c_{t}^{pv},r_{t}^{pv},k_{t}]$ be a vector of energy flow control actions at time slot $t$. Here, our objective is to minimize an average aggregated system cost consisting of: (i) the cost of energy packet transactions (selling and buying) with the P-ESP ($\overline{C}_{T_0}^{tx}$), (ii) the cost of household load scheduling delays ($C_{d}(\overline{d}_{T_{0}}^{I})$), and (iii) the cost of energy storage battery degradation ($\overline{C}_{T_0}^{s}$). Our aim is to find an optimal policy $\{\theta_{t},d_{t}^{I}\}$ while minimizing the average system cost. Thus, the problem is formulated as follows.
\begin{equation}
\stackunder{minimize}{$\{\theta_{t}, d_{t}^{I}\}$}\quad C_{d}( \overline{d}_{T_{0}}^{I})+\overline{C}_{T_0}^{tx}+\overline{C}_{T_0}^{s} \nonumber
\end{equation}
Subject to: \eqref{Eq3},\eqref{Eq4},-- \eqref{Eq10}, \eqref{Eq13}, \eqref{Eq14}, \eqref{Eq16} -- \eqref{Eq20}, and
\begin{equation}
\label{Eq30}
 r_{t}^{pv}+E_{t}^{g} \in [0, \text{min} \{H_{max}, E_{max}^{s}-E_{t}^{s} \} ] 
\end{equation}
\begin{equation}
\label{Eq31}
k_{t} \in [0, \text{min} \{K_{max}, E_{t}^{s} -E_{min}^{s}\} ]
\end{equation}
Where, $d_{t}^{I}\triangleq [d_{t}^{1},d_{t}^{2},...,d_{t}^{L}]$, and $C_{d}( \overline{d}_{T_{0}}^{I})\triangleq [C_{d}( \overline{d}_{T_{0}}^{1}),C_{d}( \overline{d}_{T_{0}}^{2}),...,C_{d}( \overline{d}_{T_{0}}^{L})]$. Clearly, the above problem is a joint stochastic optimization problem between the three considered system costs. This joint scheduling makes the problem very difficult to solve by traditional mathematical optimization techniques \cite{yang2010engineering}. Therefore, in the next section, we solve it through heuristic optimization techniques.

\section{Optimization Techniques}
\vspace{-1ex}
Heuristic algorithms are often used to solve joint stochastic optimization  due to: (i) their ability to solve high dimensional and complex problems with a fast convergence rate, (ii) ease in implementation, and (iii) capable of avoiding local optima in pursuit of a global optima \cite{yang2010engineering}.
We solve the optimization problem in Section III via three popular heuristic algorithms: genetic algorithm (GA), binary particle swarm optimization (BPSO) algorithm, and differential evolution (DE) algorithm. Their brief description is given next (more details in  \cite{yang2010engineering}).

\vspace{-1ex}
\subsection{Genetic Algorithm (GA)}
\begin{enumerate}
    \item Generate an initial population of solutions (i.e., ${\text P_{\scriptstyle{0}}}$) randomly and binary encode it such that ${\text X_{\scriptstyle \text a}} \in \{1 \:\text{if} \:{\text P_{\scriptstyle{0}}}(\text a) > 0.5, \text {otherwise} \:\: 0\}$. Each binary coded individual ${\text{X}_{\scriptstyle \text{ab}}}, \text b\in [1, \text k]$ is a k-dimensional vector denoting ON and OFF states of a given load.
   \item Use $\{E_{t}^{g},E_{t}^{pv},E_{t}^{s}, E_{t}^{i}, P_{t}^{sell}, P_{t}^{buy}\}$ as the inputs, and equations  (\ref{Eq3}),(\ref{Eq4}),(\ref{Eq6}),(\ref{Eq15}),(\ref{Eq21}) to determine the objective function in Section \ref{Problem_For}. 
   \item Determine the fitness of each individual in ${\text P_{\scriptstyle{0}}}$ with respect to the objective function in step 2 above.
   \item Adopt the process of tournament selection and select the best individuals (who perform better on objective function) from ${\text P_{\scriptstyle{0}}}$, as parents.
   \item Employ local crossover and bit-flip mutation with a probability between 0 and 1 to reproduce new individuals and update ${\text P_{\scriptstyle{0}}}$.
   \item Repeat step 2 above until the individuals in ${\text P_{\scriptstyle{0}}}$ approach the optimal values or the total number of generations reach a preset number.
    \end{enumerate} 
\vspace{-1ex}
\subsection{Binary Particle Swarm Optimization (BPSO) }
\begin{enumerate}
    \item Randomly generate an initial swarm (${\text S_{\scriptstyle{0}}}$) in a pair ($\overrightarrow {ps_i}, \overrightarrow {v_i} $), where the vector $\overrightarrow {ps_i} \in \mathrm{R^n}$ represents the position of the particles and $ \overrightarrow {v_i}$ corresponds to their velocity $ \overrightarrow {v_i} \in \mathrm{ R^n}$. Here, $\overrightarrow {ps_i}$ is computed with respect to $\overrightarrow{v_i}$ as follows.
    \begin{equation}
\label{BPSO_in}    
   \overrightarrow {ps_i}(t)=\overrightarrow{ps_i}(t-1)+\overrightarrow{v_i}(t)
\end{equation}
where $\overrightarrow {ps_i}(t-1)$ is the previous position of the particle in the swarm. 
  \item Evaluate each particle in the swarm using the input values from $\{E_{t}^{g},E_{t}^{pv},E_{t}^{s}, E_{t}^{i}, P_{t}^{sell}, P_{t}^{buy}\}$ and equations  \eqref{Eq3},\eqref{Eq4},\eqref{Eq6},\eqref{Eq15},\eqref{Eq21} to  determine  the  objective  function in Section III. If the evaluated particle minimizes the objective function then remember the particle as $p_{best}$.
 \item Update ${\text S_{\scriptstyle{0}}}$ and $\overrightarrow{v_i}$ of each particle in the swarm using \eqref{BPSO_ve}, while respecting the upper and lower bounds of $\overrightarrow{v_i}$ in \eqref{BPSO_upve}.
 \begin{equation}
\label{BPSO_ve}    
\begin{split}
\overrightarrow {v_i}(t)= \overrightarrow{v_i}(t-1)+\alpha_1 \mathrm{rand_1} \Big(p_i-\overrightarrow{ps_i}(t-1)\Big)+ \cdots \\  \alpha_2 \mathrm{rand_2} \Big(p_g-\overrightarrow{ps_i}(t-1)\Big)
\end{split}
\end{equation}
\begin{equation}
\label{BPSO_upve}
\overrightarrow {v_i}(t)=
\begin{cases}
\overrightarrow {v_i}_{\mathrm max} & \mathrm{if} \overrightarrow {v_i} > \overrightarrow {v_i}_{\mathrm max}\\
-\overrightarrow {v_i}_{\mathrm max} & \mathrm{if} \overrightarrow {v_i} < -\overrightarrow { v_i}_{\mathrm max}\\
\end{cases}
\end{equation}
In \eqref{BPSO_ve}, $\alpha_1. \mathrm{rand_1}$ and $\alpha_2. \mathrm{rand_2}$ are random weights for local and global positions ($p_i$ and $p_g$) of the particle, respectively. And $\overrightarrow {v_i}_{\mathrm max}$ and $-\overrightarrow {v_i}_{\mathrm max}$ are the maximum and minimum velocities of the particle at any point, respectively. Note that $\overrightarrow {ps_i}$ is bounded between  [0,1]. 
\item Evaluate the updated swarm (${\text S_{\scriptstyle{1}}}$) by  comparing it with ${\text S_{\scriptstyle{0}}}$ using objective function in Section \ref{Problem_For} and select the particles with lowest value of objective function and refer their position as $pg_{best}$.
\item Repeat step 4 above until particles in ${\text S_{\scriptstyle{1}}}$, ${\text S_{\scriptstyle{0}}}$ approach the optimal values or the total  number of generations reach a preset number.
\end{enumerate}

\subsection{Differential Evolution (DE)}
\begin{enumerate}
    \item Generate an initial population ${\mathrm P_e} \in \mathrm{ R^n}$ randomly using \eqref{De}, where  ${\text P_{\scriptstyle{e}}}=[p_{e1},p_{e2},p_{e3},p_{e5}, \cdots p_{en}]$.
    \begin{equation}
\label{De}
   {\text {P{\footnotesize e}}}= {p_e^L}+ {\mathrm {rand_i}}({p_e^U}-{p_e^L})
\end{equation}
    ${p_e^U},{p_e^L}$ are the upper and lower bounds of $P_{\scriptstyle{e}}$, respectively, and $\mathrm {rand_i}$ is the uniformly distributed random number between 0 and 1. Note that the individuals in ${\text P_{\scriptstyle{e}}}$ represent the operation states of appliances.  
    
    \item Generate a mutation (${M_{de}}$) vector using equation \eqref{De1}  to determine the objective function in Section \ref{Problem_For} considering values from $\{E_{t}^{g},E_{t}^{pv},E_{t}^{s}, E_{t}^{i}, P_{t}^{sell}, P_{t}^{buy}\}$ and equations (\ref{Eq3}),(\ref{Eq4}),(\ref{Eq6}),(\ref{Eq15}),(\ref{Eq21}).
  
    \begin{equation}
    \label{De1}
 {M_{de}} = v_{r1}+C(v_{r2}-v_{r3})
\end{equation}
    Where $C$ is a constant between [0,1], $v_{r1}, v_{r2},$ and $v_{r1}$ are three vectors randomly picked up from ${\text P_{\scriptstyle{e}}}$ and $r1,r2,r3$ are positive integers $\in \{1,2,3,4...n\}$.
    
    \item Generate a new trial vector $\mathrm{T_v}$ through crossover between {P{\footnotesize e}} and ${M_{de}}$ using \eqref{De2}. Calculate objective function   based on $\mathrm{T_v}$. Step (2) and (3) are compared to achieve minimal value of objective function.  
    \begin{equation}
\label{De2}
\mathrm{T_v} =
\begin{cases}
{M_{de}}& \text{if} \: \text{rand}(j) \leq \text c{\footnotesize r} \\
 \text P{\footnotesize \text e}& \text{if} \: \text{rand}(j) >\text c{\footnotesize r} \\
\end{cases}
\end{equation}
  \item Repeat step 3 above until individuals in {P{\footnotesize e}} approach the optimal values or the total number of generations reach a preset number.
\end{enumerate}

\section{ Results and discussion}\label{RD}
Considering a scheduling horizon of one day (\emph{i.e.,} 24 hours), let the PV system generate a maximum energy $E_{t,max}^{pv}$=9.62 kWh with $\eta_{pv}$=18\%, and $A_{pv}$=0.5. For simulation purpose, the solar irradiance and temperature data are taken from \cite{shirazi2015optimal}. The maximum battery storage capacity is set at 20 kW. 
In simulations, surplus energy packets are fed back to the utility via P-ESP with minimum and maximum $P_{t}^{sell}$ are 0.06 cents and 0.57 cents, respectively \cite{nojavan2017optimal}.
Similarly, energy packets are bought from utility at $P_{t}^{buy}$ which fluctuates between 0.6 cents/kWh and 3.7 cents/kWh \cite{22}.

 \begin{figure}
 \centering
 \includegraphics[width=0.9\columnwidth]{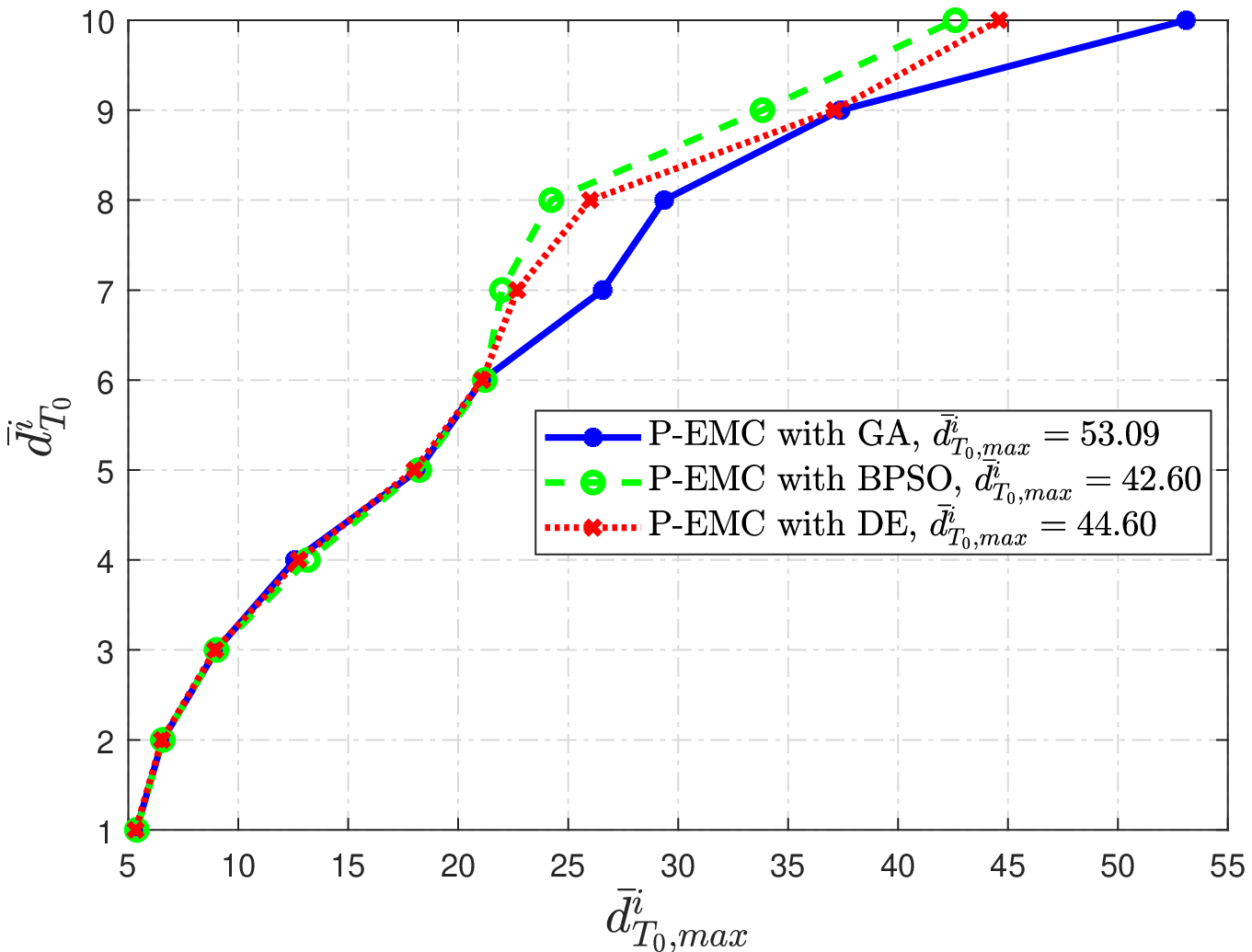}
 \caption{Delay performance of GA, BPSO, and DE: $\overline{d}_{T_0}^i$ vs $\overline{d}_{T_0, max}^i$,  $\forall i\in\{1,2,...,L\}$}
  \label{delay}
 \end{figure}
Figure \ref{delay} illustrates the relative performance between the selected algorithms (GA, BPSO and DE) in terms of average experienced load scheduling delay against the maximum allowable delay. 
It can be seen from the figure that as the maximum allowable delay requirement of loads ($\overline{d}_{T_0, max}^i$) is relaxed/increased, their average experienced delay ($\overline{d}_{T_0}^i$) also increases. However, the increase in  $\overline{d}_{T_0}^i$ is sublinear for all the compared algorithms as compared to the increase in $\overline{d}_{T_0, max}^i$.
For example, GA has achieved $\overline{d}_{T_0, max}^i$=53.09, which is greater than BPSO and DE by 11.03 \% and 19.03\% , respectively. 
%
This means scheduled load can be delayed which in turn reduces the average system cost and consequently the user QoS is compromised.

 \begin{figure}
 \centering
 \includegraphics[width=1\columnwidth]{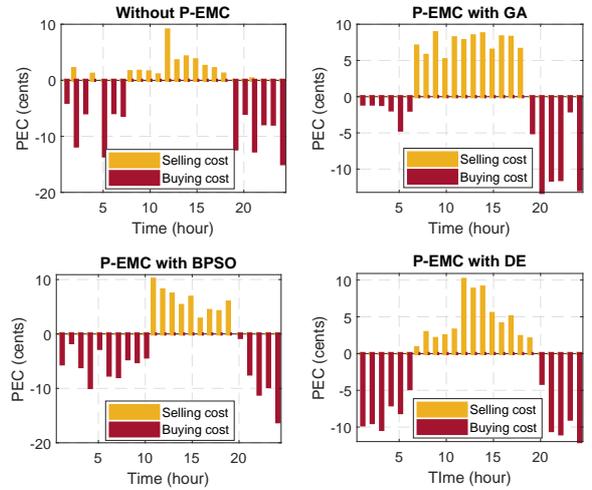}
 \caption{Performance of GA, BPSO, and DE in terms of packetized energy transactions}
  \label{PEC}
  \vspace{-2ex}
 \end{figure}

Figure \ref{PEC} depicts the relative performance of the selected algorithms (GA, BPSO and DE) employed in the P-EMC and a special case without the P-EMC in terms packetized energy transactions that involve both selling to and buying of energy packets from the utility. 
As shown in the figure, without P-EMC case has a selling cost of 34.9 cents and a buying cost of 108.08 cents in a day. 
When optimization algorithms are employed, the selling costs are increased to 88.9 cents, 54.4 cents, and 57.9 cents for P-EMC with GA, with BPSO, and with DE, respectively. Similarly, the buying costs are decreased to 67.33 cents, 99.74 cents, and 95.51 cents for P-EMC with GA, with BPSO, and with DE, respectively.
It is can be seen that GA has a higher selling cost than BPSO and DE, because GA tends to schedule the load in later time slots (achieves greater delay) and utilizes the harvested energy from the PV system and the storage system in a more efficient manner, thus it sells greater amount of energy to the utility. 
This means that the optimization algorithms help to allocate the energy resources effectively and also facilitate the user to sell back the surplus energy to the utility grid via P-ESP.
Further, BPSO and DE moderately schedule the load at time slots when energy from PV system is less or not available. Hence, achieving a relatively lower selling cost than GA.

\begin{figure}
 \centering
 \includegraphics[width=0.9\columnwidth]{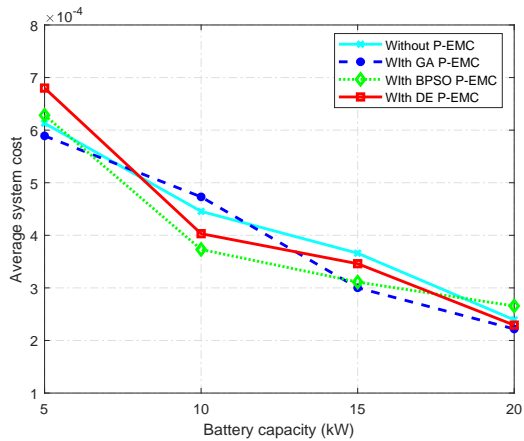}
 \caption{Average system cost vs Battery capacity}
  \label{ESS}
  \vspace{-2ex}
\end{figure}

Figure \ref{ESS} reflects the impact of battery capacity on the average system cost under the selected algorithms.
It is evident from the figure that when increase in the battery capacity induces a decrease in the average system cost for all the selected algorithms including the unscheduled special case of without P-EMC.
A higher battery capacity provides more flexibility in scheduling loads at low peak hours. Thus, resulting in reduced average system cost.
The optimization algorithms reduced average system cost to 4.7\%, 5.14\% and 1.35\% by P-EMC with GA, BPSO, and DE, respectively.  

\section{Conclusion}
This paper proposed P-EMC for a residential smart home considering household loads, energy transaction cost, PV energy generation and energy storage system. 
The proposed P-EMC employs the internal pricing model and solves the joint stochastic problem using optimization algorithms such as; GA, BPSO, and DE.
Simulation results have shown that optimization algorithms are capable to schedule the load effectively and reduced the energy procurement cost to 37.65\%, 7.5\%, and 11.5\% by GA, BPSO, and DE, respectively.
Furthermore, the proposed P-EMC helps the consumer to sell surplus energy up to 88.9  cents,  54.4  cents, and  57.9  cents with the help of GA, BPSO,  and DE, respectively.
In the future, we aim to extend our case study for the different PV generation profiles and pricing signals and analyze the level of accuracy of each designed optimization algorithm.
\section*{Acknowledgements}
\vspace{-1ex}
This work is supported by the Academy of Finland: (a) ee-IoT n.319009, (b) EnergyNet n.321265/n.328869 and (c) FIREMAN n.326270/CHIST-ERA-17-BDSI-003; and by JAES Foundation via STREAM project.

\bibliographystyle{IEEEtran}
\bibliography{Ref.bib}
\end{document}